\documentclass{JFM-FLM_Au}

\usepackage{graphicx}
\usepackage{epstopdf, epsfig}
\usepackage{amsmath,amsfonts,bm}
\usepackage{hyperref}
\usepackage{subcaption}
\usepackage{transparent}
\usepackage{color}

\lefttitle{G. D. Weymouth and M. Lauber}
\righttitle{Journal of Fluid Mechanics}

\title{Stability of Kirigami parachutes in effectively infinite numerical domains}

\corresau{Gabriel D. Weymouth, \email{G.D.Weymouth@tudelft.nl}}

\author{Gabriel D. Weymouth\aff{1} \and Marin Lauber\aff{1}}

\affiliation{\aff{1}Faculty of Mechanical Engineering (ME), Delft University of Technology, Delft, Netherlands}

\begin{document}
\maketitle
\begin{abstract}
Kirigami, the art of cutting flat sheets into deployable 3D structures, has recently inspired a new class of parachutes which can deploy into a naturally stable inverted canopy. However, the dynamic mechanism, fluid forces, and geometrical parameters that grant this stability have not yet been clearly identified.
In this paper, we use a novel Biot-Savart far-field boundary condition to perform prescribed acceleration and free-falling simulations in effectively infinite domains, tracking the descent of a parameterized kirigami parachute. The far-field velocity is reconstructed from the interior vorticity, resulting in less than 0.1\% variation in the predicted dynamics as the domain size is doubled.
We first show the linear forces drop $2-5\times$ as the parachute is deployed due to increased permeability, whereas the moments increase due the counterbalancing effect of the increased lever-arm.
Next, we find that the kirigami parachute achieves stable flight for deployment heights as small as half its radius, quickly damping out applied perturbations. 
For smaller deployments, the parachute tumbles due to side-slip and rotational coupling, as in falling disks.
These effectively unbounded simulations identify that deployments approximately equal to the radius offer high drag forces with strong dynamic stability, providing a simple design rule for deployable parachutes. 

\end{abstract}


\section{Introduction}\label{sec:intro}

The flight dynamics and stability of falling objects depends on the interplay of the geometric properties and fluid forces. The falling dynamics of thin plates and coins have been well studied \citep{Andersen2005UnsteadyPlates, Auguste2013FallingDisks}, exhibiting tumbling, flutter, wandering, and chaotic transitions between these modes, with a strong dependence on the object's thickness and inertia. Introducing porosity, such adding a hole to coins \citep{Vincent2016HolesCoins} acts to disrupt the wake and fluid forcing, stabilize falling, while very high porosity fallers such as dandelion seeds \citep{Cummins2018ADandelion,Bose2025PorousIncidence} exhibit passively stable descent.

Kirigami, the art of cutting flat sheets into deployable 3D structures, has introduced a new class of falling porous structures with the development of kirigami parachutes that extend passively under the tension between fluid drag and their payload.
The fluid-structure interaction of kirigami reconfiguration into stable parachute-like shapes was studied experimentally first by \cite{Carleton2024KirigamiFlow} and the free-falling motion of this class of parachutes demonstrates passively stable decent in experiments \citep{Lamoureux2025Kirigami-inspiredReconfiguration}. However, the dynamic stability mechanism of the deployed parachute and its sensitivity to geometric features and fluid reaction forces have not yet been clearly identified.

Using numerical simulations to study this problem in detail poses a significant conflict between the small-scale geometric and vortical flow features of the porous body and its turbulent wake and the large-scale (even unbounded) domains required to establish the long-time falling dynamics.
Although Eulerian immersed boundary methods can efficiently capture the complex deployed geometry and motion of the kirigami using fine Cartesian meshes, fixed-domain Eulerian methods suffer from significant blockage errors unless the computational domain is tens or even hundreds of body lengths in size \citep{Lauber2022,Bose2025PorousIncidence}. In contrast, vortex-based Lagrangian methods naturally satisfy far-field boundary conditions but must generate fresh persistent vortex-elements from every kirigami edge at every time step, requiring special methods for long-time simulations \citep{Koumoutsakos1995High-resolutionMethods,Gillis2019AMethod}.

In this paper, we enable long-time free-fall simulations in effectively infinite domains by merging the two approaches, developing far-field boundary conditions for an Eulerian flow solver which are based on the Biot-Savart integral used in vortex methods. We apply this approach to simulate the unsteady flow around parametrically defined kirigami parachutes, identify the side-slip flutter mechanism governing their stability, and derive a simple geometric design rule for stable deployment.

\section{Geometric Description and Numerical Method}\label{sec:method}

\begin{figure}
    \centering
    \includegraphics[width=\textwidth]{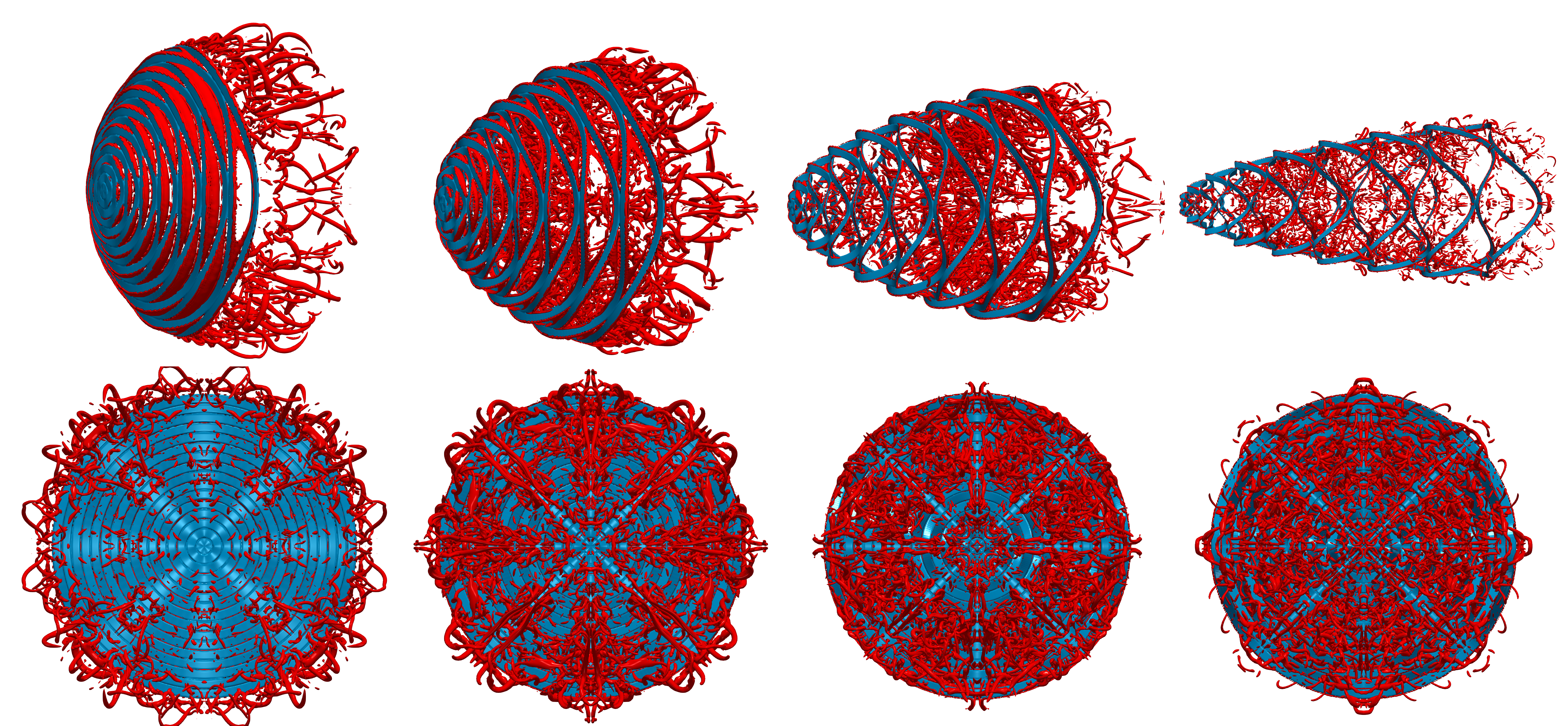}
    \caption{Isometric and downstream views of the kirigami parachutes with different deployment ratios, increasing from left to right in $H\in[0.5,1.0,2.0,4.0]$. The unsteady flow structures are visualized using isosurfaces of the $\lambda_2$-criterion with $\lambda_2L^2/U^2=-0.02$ at time $tU/L=3$.}
    \label{fig:kirigami}
\end{figure}

\noindent Our idealized kirigami parachute consists of a set of $N_r$ concentric rings of vanishing thickness and equal width, filling a circular disk of radius $R$. Adjacent rings are connected at only four places along the azimuthal direction, allowing them to separate between these points, Fig.~\ref{fig:kirigami}. This work focuses on the fluid mechanical and flight stability problems, neglecting the torsional deformation of paper kirigami in favor of a prescribed deployment height-to-radius ratio $H$ and a parabolic profile.

\begin{figure}
    \centering
    \vspace{2mm}
    \def\svgwidth{0.6\columnwidth}
    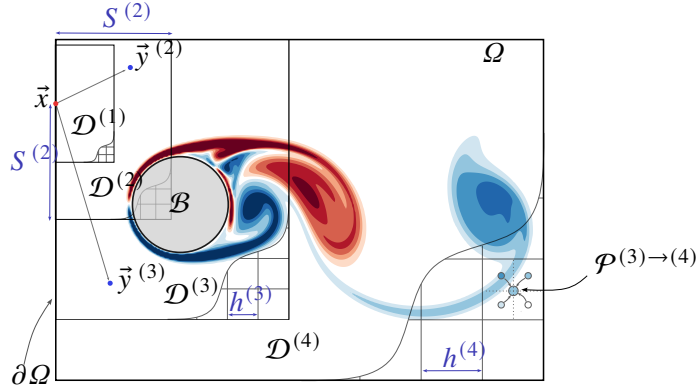
    \caption{Schematic of the computational domain $\Omega$, its boundary $\partial\Omega$, and the immersed body $\mathcal{B}$. The figure also shows the multilevel evaluation of the Biot-Savart integral for a boundary point $\bm x$ over the nested subdomains $\mathcal{D}^{(1)} \cup \mathcal{D}^{(2)} \cup \mathcal{D}^{(3)} \cup \mathcal{D}^{(4)}=\Omega $. The $l=2$ domain half-width is shown, ${S}^{(2)}$, together with a schematic of the coarsening of the grid between the domains and the multilevel pooling $\mathcal{P}^{(3)\to(4)}$.}
    \label{Fig_2}
\end{figure}

The kinematics of the free-falling kirigami parachute are obtained by solving the three degrees of freedom equations of motion in the x-y plane:
\begin{equation}\label{eq:motion}
    \begin{split}
        m\frac{\text{d}\bm{u}_\mathcal{B}}{\text{d}t} = (\mathbb{I}-\hat{e}_3)\cdot\bm{f}_\text{fluid} + mg\cdot\hat{e}_1,\quad
        \bm{f}_\text{fluid}=\oint_{\partial\mathcal{B}}\bm \sigma\text{ d}S \\
        I\frac{\text{d}^2\bm{\theta}_\mathcal{B}}{\text{d}t^2} = \hat{e}_3\cdot\bm{\tau}_\text{fluid},\quad
        \bm{\tau}_\text{fluid}=\oint_{\partial\mathcal{B}}(\bm{x}-\bm{x}_c)\times\bm\sigma\text{ d}S
    \end{split}
\end{equation}
where $\bm{u}_\mathcal{B}$ and $\bm{\theta}_\mathcal{B}$ are the body velocity and rotation angle, while $m$, $\bm x_c$, and $I$ are the mass, center of mass, and rotational inertia of the kirigami parachute, and $g$ is gravitational acceleration.
The fluid force and moment are given by integrals of the fluid traction
\begin{equation}
    \bm\sigma = \left[-p\mathbb{I} + \rho \nu\left(\nabla\bm u + \nabla\bm u^T\right)\right]\cdot\bm{n}
\end{equation}
over the body boundary $\partial \mathcal{B}$, where the pressure $p$ and velocity $\bm u$ are governed by the incompressible unsteady Navier-Stokes equations
\begin{equation}\label{eq:navier-strokes}
     \frac{\partial\bm{u}}{\partial t} +\left(\bm{u}\cdot\nabla\right)\bm{u} = -\frac 1 \rho \nabla p + \nu\nabla^2\bm{u}, \quad
    \nabla\cdot\bm{u}=0, \quad\quad\forall\ \bm{x}\in\Omega,
\end{equation}
and where $\rho,\nu$ are the fluid's density and viscosity and $\Omega$ is the fluid domain, see Fig.~\ref{Fig_2}. 

These equations are supplemented with the no-slip body boundary condition
\begin{equation}
    \bm{u} = \bm{u}_\mathcal{B} \quad \forall\ \bm{x} \in \partial\cal{B}
\end{equation} 
where $\bm{u}_\mathcal{B}$ is obtained by integrating the 
equations of motion (Eq.~\ref{eq:motion}) with an explicit Verlet scheme. 
To stabilize the explicit fluid-structure coupling, the added-mass tensor is computed for each geometry and subtracted from both sides of the momentum equation during integration. This removes the instability that otherwise arises when the body and fluid inertia are comparable without requiring iterative coupling \citep{Weymouth2014ChaoticCylinder}. The resulting accelerations and velocities define an accelerating reference frame that tracks the descending parachute 
(see supplementary videos).

The flow is also subject to a velocity condition on the domain boundary $\partial\Omega$. 
Assuming the vorticity generated by the immersed body is confined within the computational domain, we can replace the typical ``far-field'' boundary conditions by a Biot-Savart integral over the vorticity $\bm \omega$ inside the domain
\begin{equation}\label{eq:Biot}
    {\bm u}({\bm x}) = \bm U_\infty+f_n(\bm x; \bm\omega,\Omega) = \bm U_\infty+\int_\Omega K_{n}({\bm x} - \bm{y})\times \bm\omega({\bm y})\text{ d}\bm{y} \quad\quad \forall\, \bm x\, \in \partial\Omega
\end{equation}
where $\bm U_\infty$ is the domain velocity and $f_n$ is the convolution of $\bm\omega$ with the $n$-dimensional Biot-Savart kernel $K_2(\bm r)= -\frac{\bm r}{2\pi|r|^2}$, $K_3(\bm r)= -\frac{\bm r}{4\pi|r|^3}$.
Two issues must be addressed to use Eq.~\ref{eq:Biot} as the far-field condition within an incompressible Eulerian solver: (i) the pressure and domain velocity at each time step must be determined simultaneously, and (ii) the convolution $f_n$ must be calculated efficiently.

(i) To illustrate the pressure-velocity coupling, consider the first step of the standard pressure projection scheme over a time step $\Delta t$
\begin{align}\label{eq:update}
    \bm{u}^{t+\Delta t}=\bm{u}^*-\frac {\Delta t}\rho\nabla p \quad\forall\ \bm{x}\in\Omega,\quad\text{where}\quad  \bm{u}^* = \bm{u}\,^t + \Delta t\left[\nu\nabla^2\bm{u} -\left(\bm{u}\cdot\nabla\right)\bm{u}\right]
\end{align}
is the intermediate velocity.
Substitution of Eq.~\ref{eq:update} into Eq.~\ref{eq:Biot} gives
\begin{align}\label{eq:biot-press}
    &\bm{u}^{t+\Delta t}(\bm x) = f_n(\bm x,\nabla\times\bm{u}^*,\Omega)-f_n\left(\bm x,\nabla\times\frac{\Delta t}{\rho}\nabla p,\Omega\right) \quad\forall\ \bm x \in \partial\Omega
\end{align}
While $\nabla\times\nabla p=0$ in the domain interior, the pressure gradient generates a thin layer of vorticity on the body boundary $\partial\cal B$ every time step \citep{Morton1984GeophysicalTy} which modifies the domain velocity via Eq.~\ref{eq:biot-press}.
We therefore must solve for the two simultaneously. In this work, we use a multigrid solver for the pressure Poisson system, as described in \cite{Weymouth2022Data-drivenProjection}, and update the domain velocities and corresponding Poisson residuals with Eq.~\ref{eq:biot-press} at the end of each V-cycle. In practice, we find updating the domain velocity within the pressure loop never increases the required number of iterations to reach convergence, and indeed typically decreases the required iterations on small domains compared to standard domain BCs.



(ii) A naive evaluation of the integral $f_n$ 
would require computing the interaction of all $N_s$ vortex sources in the domain $\Omega$ with all $N_t\sim O(N_s^{(n-1)/n})$ velocity targets on the domain boundary $\partial\Omega$. The 3D Cartesian-grid simulations used in this work have $N_s \sim 10^8$ and $N_t\sim 2.5\times 10^5$, making direct evaluation prohibitively expensive. 
We therefore develop a Fast Multi-level Method (FM$\ell$M) that exploits the Cartesian grid to evaluate all the target velocities in $O(N_s)$ operations, detailed in Appendix~\ref{app:validation}. As illustrated in Fig.~\ref{Fig_2}, FM$\ell$M achieves this speed-up by constructing a multi-level vorticity field by recursively pooling, partitioning the domain into nested subdomains around all the multi-level boundary targets, and then accumulating the induced velocity from coarse to fine levels. 

Appendix~\ref{app:validation} validates the Biot-Savart far-field conditions, recovering the correct near body flow and forces for canonical circle and sphere cases using domains only a few body lengths long; more than 100x smaller than the domain required by reflection BCs. 
The fundamental assumption of this approach is that the flow external to the domain is irrotational, but the results are remarkably invariant to domain size even when the vortical wake is truncated by the domain boundary. 
We also find the new far-field conditions are extremely efficient.
We use WaterLily.jl, an immersed-boundary flow solver based on the Boundary Data Immersion Method (BDIM) \citep{Maertens2015} that can execute any backend \citep{Weymouth2025WaterLily.jl:Bodies} for all simulations, and measure that only around 5\% of the simulation time is spent in the FM$\ell$M Biot-Savart update. The complete code and instructions to reproduce these results are available online
\footnote{\url{https://github.com/weymouth/BiotSavartBCs.jl}}.

\section{Results}\label{sec:results}

We first simulate the flow and dynamics of the kirigami parachute using prescribed impulsive acceleration $\bm{u}_\mathcal{B}(t) = [\min(U,at), 0, 0]^\top$, where $a=R/2U^2$, and a final Reynolds number of $\text{Re}=\frac{U2R}\nu=10^4$. This prescribed velocity profile produces an increasing drag force during the acceleration phase, followed by relaxation to the steady drag force regardless of the parachute shape. We use a grid resolution of 85 cells per $R$ and set the ring thickness to only 3 cells, or $0.035R$. The computational domain is set to $L\times3R\times3R$, where $L=5R+H$ to accommodate different deployment lengths. Flow fields at the end of the acceleration phase are shown in Fig.~\ref{fig:kirigami}.

\begin{figure} 
    \centering 
    \includegraphics[width=\textwidth]{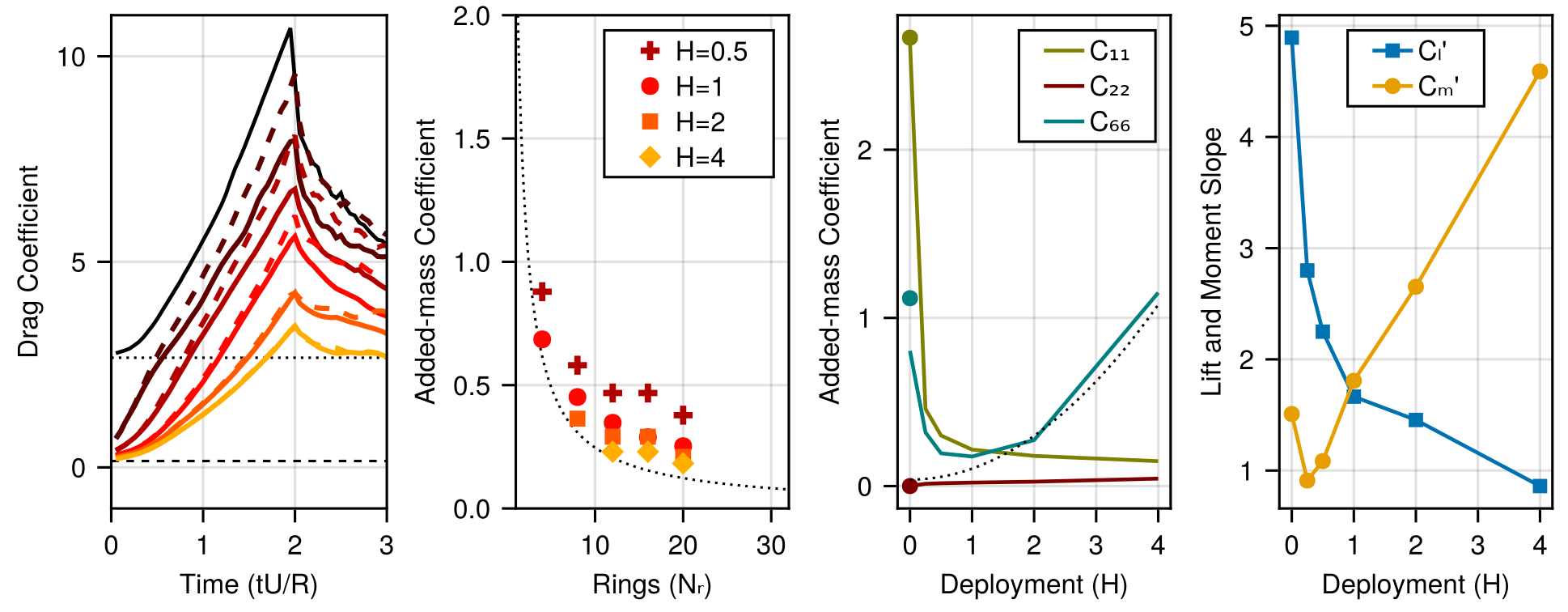} \caption{(Left) Drag coefficient histories for deployment ratios $H$. The horizontal lines are the theoretical added-mass coefficient of a solid disk ($H=0$, dotted) and the ring system Eq.~\ref{eq:C11_C66} ($H\to\infty$, dashed). (Center-left) Variation of the initial added mass of a parachute with deployment $H=1$ for different numbers of rings $N_r$ and theoretical $C_{11}$ limit (dotted). (Center-right) Measured linear and rotational added-mass coefficient for different parachute deployment $H$ with $N_r=16$ and theoretical $C_{66}$ limit (dotted). (Right) Lift and moment curve slope against deployment height $H$ for parachutes with $N_r=16$.} \label{fig:Cd}
\end{figure}

We parametrically vary the deployment ratio $H \in 0\ldots 4$, the number of rings $N_r \in 4\ldots20$, and use two static drop angles $\theta \in \{0,0.2\}$rad relative to gravity. The resulting forces and moments, scaled by $\frac 12 \rho U^2 R^2$ and $\frac 12 \rho U^2 R^3$ respectively, are shown in Fig.~\ref{fig:Cd}. Increasing the deployment $H$ has a significant effect on the force, with the increased permeability reducing the steady drag by around a factor of two by $H=4$ compared to $H=0$ (a solid disk). The peak drag is even more sensitive to $H$, dropping more than $3\times$ because of the large added-mass contribution during acceleration. The side force per-radian $C_L'$ drops by more than $5\times$ due to permeability over the full range of $H$ studied, whereas the restoring moment per radian around the nose $C_M'$ displays competing effects; dropping initially due to permeability, but then increasing by $5\times$ due to the increased lever arm provided by the deployment.

In contrast, we find that increasing $N_r$ only asymptotically decreases the drag for a given deployment $H$, with little change beyond $N_r>12$. We can model this asymptotic behavior and develop a lower bound for the added mass coefficients of the kirigami structure using potential flow theory. The added mass of ring $i$ can be approximated as that of a 2D plate times the circumference, and the added inertia is half that value times the radius squared.
\begin{equation}
    m_{11,i} \approx \rho \pi\left(\frac w 2\right)^2 2\pi \frac{r_i+r_{i-1}}2\,,\quad  m_{66,i} \approx \rho \pi\left(\frac w 2\right)^2 \pi \left(\frac{r_i+r_{i-1}}2\right)^3
\end{equation}
where $w = R/N_r$ and $r_i=iw$, are the width and outer radius of each ring. 
The $m_{22}$ value is much smaller, but nonzero due to the finite ring thickness.
A lower bound on the added mass coefficients, neglecting the interaction of the rings,  is the sum of the ring contributions normalized by $\rho R^3$ and $\rho R^5$ respectively
\begin{equation}\label{eq:C11_C66}
    C_{11} \ge \frac{\pi^2}{4N_r}\,,\quad C_{66} \ge \frac {\pi^2} {16N_r} + \beta N_r H^2
\end{equation}
where we have neglected the small $1/N_r^3$ contribution to $C_{66}$ but included the factor proportional to $m_{22}H^2$ from the parallel axis theorem for each ring.

Figure~\ref{fig:Cd} shows the added-mass coefficients as measured in the first time step of the acceleration phase. As predicted by this simple model, the added mass decreases inversely with the number of rings and approaches the lower bound for large $H$. The measured added inertia $C_{66}$ demonstrates the same competition between permeability and lever-arm as $C_M'$, with a minima around $H=1$.

We next investigate the flow and dynamics of the kirigami parachute under free-fall.
The kirigami parachute is initially at rest in a quiescent fluid, with an initial parachute angle $\theta_{0}$ relative to gravity.
Once released, the parachute is accelerated downward following Eq.~\ref{eq:motion}. To mimic a parachute under load, we concentrate the mass $m=0.84\rho R^3$ at the nose and set $I \ll \rho R^5$, meaning the added fluid inertia dominates. Defining a nominal drop rate $U=\sqrt{gR}$ for nondimensionalization, the Reynolds number matches the prescribed tests above, as does the grid resolution and ring thickness.

\begin{figure}
    \centering
    \includegraphics[width=\linewidth]{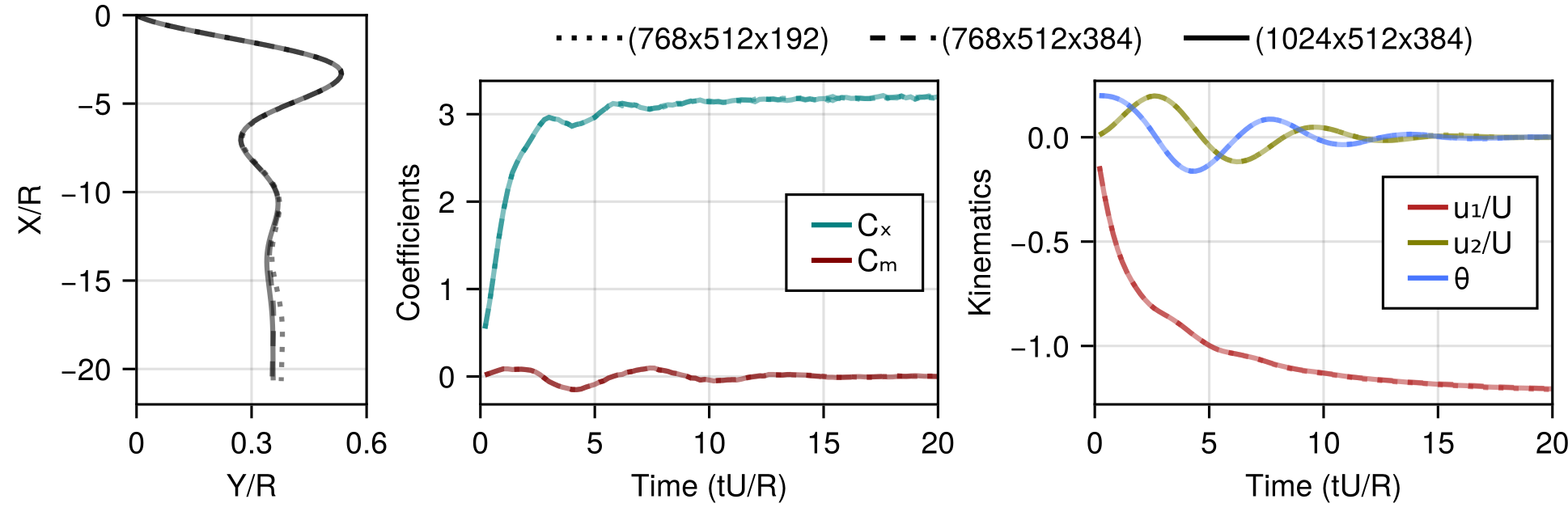}
    \caption{Sensitivity of the kirigami parachute dynamics to the domain size. (Left) Trace, (center) force and moment coefficients, and (right) velocity and rotation angle during fall for different domains of $75$M, $150$M, and $200$M cells, corresponding to the dotted, dashed and solid line, respectively.}
    \label{fig:kirigami_domain}
\end{figure}

Fig.~\ref{fig:kirigami_domain} shows the dynamics for the perturbed kirigami parachute using $H=1$ and $N_r=16$, which match the strongly stable dynamics measured in the experiments of \cite{Lamoureux2025Kirigami-inspiredReconfiguration}. The results also confirm that the new Biot-Savart Boundary far-field conditions enable dynamic predictions with negligible domain size dependence.
The resulting force and moment coefficients, scaled body velocities, and angle are all visually indistinguishable. Even the integrated trajectory deviates by less than $0.02R$ after dropping $20R$ as the domain area is reduced by nearly a factor of three.

\begin{figure}
    \centering
    \includegraphics[width=\linewidth]{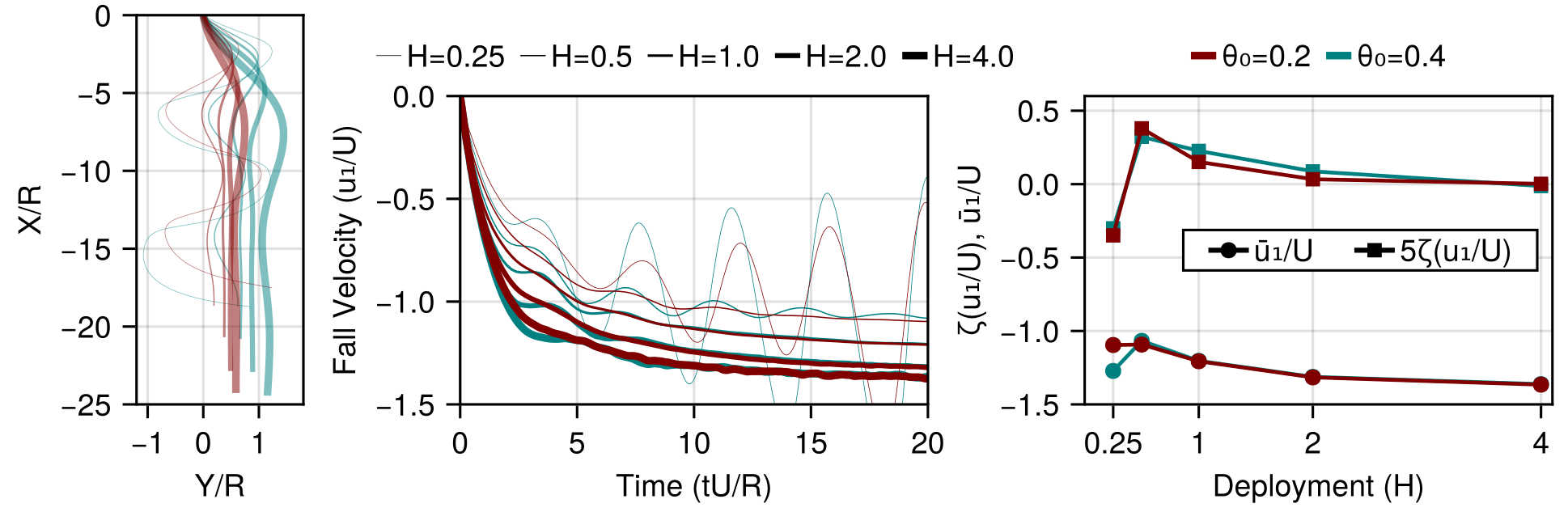}
    \caption{(Left) falling path, (center) vertical falling velocity and (right) the damping coefficient and terminal falling velocity of the kirigami parachute for different initial rotation angle $\theta_0$ (line color) and deployment ratio $H$ (line thickness). The damping coefficient is obtained from a log-decrement test of $u_{{\cal B},1}$ and the terminal falling velocity is the average of the last oscillation cycle.}
    \label{fig:results}
\end{figure}

We next study the impact of deployment length on the falling dynamics of the kirigami parachute. Fig.~\ref{fig:results} summarizes the free-falling trajectories with deployment lengths $H\in 1/4 \ldots 4$ and starting angles $\theta_0 \in \{0.2, 0.4\}$rad to investigate their stability. As deployment increases, the terminal falling velocity also increases due to the reduced drag coefficient. While a reduced drop rate is beneficial for a parachute, stability is a critical constraint, and the low-deployment parachutes with $H<1/2$ demonstrate unstable growth of their initial perturbation. Since these parachutes rotate rapidly, we stop the simulations once tumbling starts, when the rotation angle, forces, and moment peak.

This instability is due to side-slip and pitch flutter. Scaling all velocities by the nominal drop rate $U=\sqrt{gR}$, lengths by $R$, and time by $R/U$, the dimensionless perturbation equations are
\begin{equation}
    m^* \dot v_{\cal B} +\tfrac 12 C_L'\,(\theta-v_{\cal B})=0, \qquad
    I^*\ddot\theta + C_\Omega\,\dot\theta
    + \tfrac 12C_M'\,(\theta-v_{\cal B}) = 0,
\end{equation}
where $m^* = (m+m_{22})/\rho R^3$ and $I^* = (I+m_{66})/\rho R^5$ are the scaled effective linear and rotational inertia, $v_{\cal B}=u_{{\cal B},2}/U$ is the scaled horizontal velocity making $\theta-v_{\cal B}$ the flow angle relative to the parachute centerline, $C_\Omega$ is the rotational damping, and $C_L'$, $C_M'$ are the side-force and moment derivatives with respect to angle. We have neglected the cross-coupling added mass $m_{26}$ and any depth $d$ of the payload center below the parachute nose; both add static stability but don't clarify the observed flutter mechanism. 

Substituting perturbations of the form $v_{\cal B},\theta\sim e^{i\omega_n t}$ and combining the equations gives
\begin{equation}
    \ddot\theta + 2\zeta\omega_n\,\dot\theta+\omega_n^2\,\theta = 0 \quad\text{with}\quad 2\zeta\omega_n=\frac{C_\Omega}{I^*}- \frac{C_L'}{2m^*}\,,\quad \omega_n^2=\frac{C_M'm^*-C_L'C_\Omega}{2I^*m^*}
\end{equation}
meaning the coupled side-slip and rotation will lead to a flutter instability when
\begin{equation}
    C_L'\,I^* > 2C_\Omega\, m^*. \label{eq:flutter}
\end{equation}
Equation~\ref{eq:flutter} identifies the destabilizing role of side-force generation and rotational added inertia, which are substantial for $H\le1/4$ but drop rapidly with $H$, Fig~\ref{fig:Cd}. In contrast, $m^*\approx m/\rho R^3$ for all $H$, and we expect $C_\Omega$ to quickly stabilize due to the moment-arm effect, similar to $C_M$, such that the balance changes quickly to favor stable falling. The log-decrement measurements of the effective damping jump from $\zeta(H=1/4)=-0.02$ to $\zeta(H=1/2)=0.02$, Figure~\ref{fig:results}. Further increasing the deployment, the parachute remains stable, but the damping coefficient gradually decreases due to the increased inertia for $H>1$.


\section{Conclusion}\label{sec:conclusion}

This work develops far-field boundary conditions for Eulerian simulations based on the Biot-Savart integral and applies them to study the falling dynamics of parametrically defined kirigami parachutes. The optimized Biot-Savart method adds only around 5\% computational cost to Eulerian solvers with less than 0.1\% variation in predicted dynamics even as the vortical wake is truncated by the domain boundary, enabling effectively unbounded simulations without significant computational cost.

We use this approach to study the deployment-to-radius ratio $H$ as the governing parameter for kirigami parachute dynamics and identify the key physical competition at play; increasing $H$ strongly reduces the local stresses due to the increased permeability and wake disruption but also increases the vertical lever-arm between the stresses. This interplay rapidly shifts the side-slip rotation flutter dynamics from unstable to stable as the kirigami deploys, explaining the passively stable falling trajectories observed in \cite{Lamoureux2025Kirigami-inspiredReconfiguration}. 

While the present results assume a prescribed deployment shape, neglecting reconfiguration dynamics, the identified stability mechanism depends only on the deployed geometry rather than the path to it. A deployment ratio of $H=\tfrac{1}{2}$ offers the highest deceleration at the stability boundary for this geometry and payload model, and the simple dynamic stability criterion developed provides a framework to study general statically stable falling systems.

\appendix
\section{Fast Multi-\textit{level} Method and Validation Cases}\label{app:validation}

This appendix details the Fast Multi-\textit{level} Method (FM$\ell$M) used to evaluate the Biot-Savart integral Eq.~\ref{eq:Biot} with $O(N_s)$ operations, where $N_s$ is the number of vorticity sources, and validates the resulting far-field boundary conditions against canonical 2D and 3D flows.

Fast Multipole Methods (FMMs) approximate $N \times N$ interactions of an arbitrarily-spaced point-cloud to $O(N\log N)$ or $O(N)$ \citep{Greencard1987ASimulations, Fong2009TheMethod}, but our application allows for significant simplifications. First, the uniform Cartesian-grid enables a purely geometric multilevel decomposition, while the finite-volume representation enables a simple pooling operation to conserve total circulation to machine precision, as recommended by \cite{Colonius2008} and others. Second, since Eq.~\ref{eq:Biot} is only applied to domain boundary velocities and not field points, truncating the decomposition has no impact on the satisfaction of the governing equations~\ref{eq:navier-strokes}. 

Based on this, we construct a multi-level vorticity field which leverages the uniform source distribution. We denote this field $\omega^{(i)}$ for $i\in 1\ldots l$ where level $i=1$ is the finest and $l$ is the coarsest and the cell size doubles with each level, $h^{(i+1)}=2h^{(i)}$. The number of sources drops geometrically at each level, $N_s^{(i+1)}=N_s^{(i)}/2^n$, giving less than $\frac{2^n}{2^n-1} N_s$ sources overall. The levels are filled from finest to coarsest recursively 
\begin{equation}
    \omega^{(i+1)}=\mathcal{P}^{(i)\to(i+1)}\omega^{(i)}
\end{equation}
where the pooling operator $\mathcal{P}$ for a cell at level $i+1$ sums the circulation $\Gamma=\omega h^2$ of the corresponding sub-cells in level $i$, Fig.~\ref{Fig_2}. Filling the multi-level field requires $O(N_s)$ operations and is trivially parallel over the sources on each level.

We use this field to efficiently approximate the integral Eq.~\ref{eq:Biot} by partitioning the domain $\Omega$ into nested subdomains $\mathcal{D}^{(i)}(\bm x)$ around point $\bm x$, Fig.~\ref{Fig_2}. Each domain has a half-width $S^{(i)}=h^{(i)}\tilde S_n$, where $\tilde S_n$ is a threshold distance that depends only on the flow dimension. Defining the multi-level set of targets $\bm  x^{\,(i)}$ on the boundaries of each level, we can then compute the level-$i$ velocity induced by vortex interactions 
\begin{equation}\label{eq:interaction}
    \bm v^{\,(i)}(\bm x^{\,(i)}) = f(\bm x^{\,(i)}; \bm \omega^{(i)},\mathcal{D}^{(i)}(\bm x^{\,(i)})).
\end{equation}
using $O(\tilde S_n^n)$ operations. The velocity target count $N_t$ also drops geometrically across levels, so there are $O(N_t)$ multi-level velocities, all of which can be computed independently in parallel.
Since $\bigcup_i {\cal D}^{(i)}(\bm x)= \Omega$, the total induced velocity is the sum of the interaction velocities at each level. These are recursively accumulated from the coarsest level to the finest
\begin{equation}\label{eq:unpool}
    \bm u^{\,(i)} \approx \bm v^{\,(i)}+\mathcal{P}^{(i+1)\to(i)} \bm u^{\,(i+1)}
\end{equation}
where $u^{\,(l+1)}=0$ and the unpooling operator is bi-linear interpolation from locations $\bm x^{\,(i+1)}$ to  $\bm x^{\,(i)}$. Unpooling requires $O(N_t)$ operation, making $O(N_s)$ the dominant scaling for our FM$\ell$M.

We validate the FM$\ell$M Biot-Savart far-field condition on two canonical cases: the flow induced around a 2D circular cylinder, where the entire wake remains within the domain, and a 3D sphere allowed to evolve until the wake is truncated by the domain boundary, directly testing the irrotationality assumption of Eq.~\ref{eq:Biot}.

\begin{figure}
    \centering
    \includegraphics[width=\textwidth]{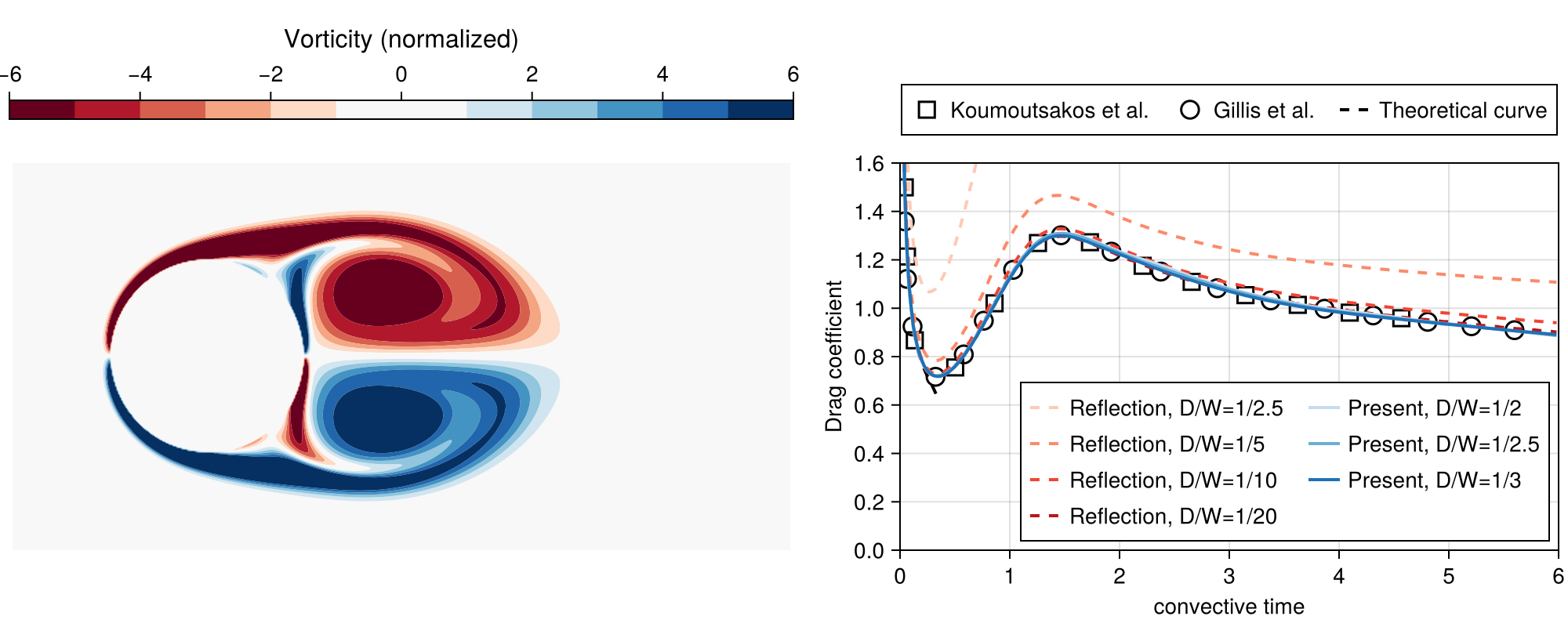}
    \caption{Impulsively started circular cylinder at $\text{Re}=550$ validation case. (Left) Vorticity field $\omega D/U$ at convective time $tU/D=4$ over the full $2D\times 4D$ domain. (Right) Early time history of the drag coefficient $C_d=F_d/\frac 12 \rho U^2 D$ for various domain blockage ratios $D/W$. The present Biot-Savart BCs match the theoretical and vortex-based methods even with 50\% blockage.}
    \label{fig:cylinder_force}
\end{figure}

First, a circular cylinder of diameter $D$ is immersed in a viscous fluid and impulsively accelerated to free-stream velocity $U$, reaching a final Reynolds number of $\text{Re}=\frac{UD}{\nu}=550$. We set $D=128$ cells for all tests and use a $2\times 1$ rectangular domain, varying the width $W$ from $2D$ to $20D$. We use a FM$\ell$M subdomain half-width of $\tilde S_2=4$ for 2D flows. Fig.~\ref{fig:cylinder_force} (left) shows the vorticity field on the smallest domain at convective time $t^*=tU/D=4$.

Fig.~\ref{fig:cylinder_force} -- right compares the predicted drag coefficient using the new Biot-Savart boundary conditions to the analytical solution valid for for $t^*\ll 1$ \citep{Collins1973TheCylinder} and two sets of reference data from high-resolution vortex-based methods \citep{Koumoutsakos1995High-resolutionMethods, Gillis2019AMethod}. The new condition is able to accurately predict the total drag force on the cylinder, matching the theoretical and vortex-based methods even when using the smallest $4D\times2D$ domain with 50\% blockage. 
In contrast, reflective boundary conditions induce extremely large blockage effects for small domains and only slowly converge to the external flow solution. Even using a domain 100 times larger produces noticeable deviations for $t^*>3.5$.

\begin{figure}
    \centering
    \begin{subfigure}{1.0\textwidth}
        \centering
        \includegraphics[width=\textwidth]{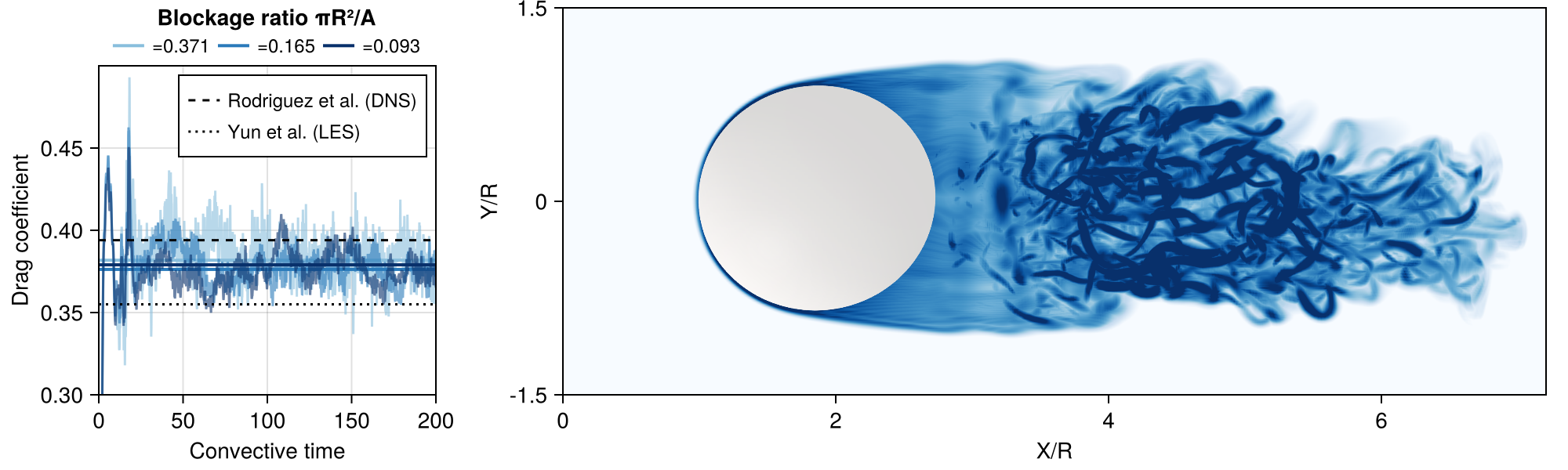}
    \end{subfigure}
    \caption{Sphere flow vortex wake and drag coefficient at $Re = 7400$ with a resolution of 44 cells per radius. (Left) Instantaneous and time-averaged drag coefficients for three domains, where $A$ is the domain frontal area. The time-averaged drag is obtained by averaging the last 100 convective times.
    (Right) Wake visualised by using $|\omega|R/U$ at time $tU/R=53$ for the entire $3.6D\times1.5D\times1.5D$ domain.}
    \label{fig:sphere}
\end{figure}

Second, a sphere of radius $R$ is immersed in a viscous incompressible flow and the flow is impulsively started with a uniform velocity $U$. The Reynolds number is set to $Re=U2R/\nu=7400$, we use a resolution of 44 cells per radius and use a domain size of $2.5C \times C \times C$ cells varied over $C\in[128,192,256]$. The wake extends outside the computational domain within 12.5 convective time units, and we perform the simulations until convective time $tU/R=200$. The faster decay of the 3D Biot-Savart kernel allows us to use a subdomain half-width of only $\tilde S_3=2$ for 3D simulations. 

Fig.~\ref{fig:sphere} shows the drag coefficient when using the Biot-Savart boundary condition is insensitive to the domain size - even for domains only 50\% wider than the sphere diameter. The averaged drag over the last 100 convective time units is $C_D=0.38 \pm 0.004$ for all blockage ratios, which compares favorably to the DNS and LES simulation results obtained in \citep{Yun2006VorticalNumbers, Rodriguez2011Direct3700}. 

\bibliographystyle{jfm}
\bibliography{references}

@article{Colonius2008,
    title = {{A fast immersed boundary method using a nullspace approach and multi-domain far-field boundary conditions}},
    year = {2008},
    journal = {Computer Methods in Applied Mechanics and Engineering},
    author = {Colonius, Tim and Taira, Kunihiko},
    number = {25-28},
    pages = {2131--2146},
    volume = {197},
    doi = {10.1016/j.cma.2007.08.014},
    issn = {00457825}
}

@article{Maertens2015,
    title = {{Accurate Cartesian-grid simulations of near-body flows at intermediate Reynolds numbers}},
    year = {2015},
    journal = {Computer Methods in Applied Mechanics and Engineering},
    author = {Maertens, Audrey P. and Weymouth, Gabriel D.},
    pages = {106--129},
    volume = {283},
    publisher = {Elsevier B.V.},
    url = {http://dx.doi.org/10.1016/j.cma.2014.09.007},
    doi = {10.1016/j.cma.2014.09.007},
    issn = {00457825}
}

@article{Lauber2022,
    title = {{Immersed boundary simulations of flows driven by moving thin membranes}},
    year = {2022},
    journal = {Journal of Computational Physics},
    author = {Lauber, Marin and Weymouth, Gabriel D. and Limbert, Georges},
    month = {5},
    pages = {111076},
    volume = {457},
    publisher = {Academic Press},
    doi = {10.1016/J.JCP.2022.111076},
    issn = {0021-9991},
    arxivId = {2110.06535}
}

@article{Gillis2019AMethod,
    title = {{A 2D immersed interface Vortex Particle-Mesh method}},
    year = {2019},
    journal = {Journal of Computational Physics},
    author = {Gillis, T. and Marichal, Y. and Winckelmans, G. and Chatelain, P.},
    month = {10},
    pages = {700--718},
    volume = {394},
    publisher = {Academic Press},
    doi = {10.1016/J.JCP.2019.05.033},
    issn = {0021-9991}
}

@article{Greencard1987ASimulations,
    title = {{A Fast Algorithm for Particle Simulations}},
    year = {1987},
    journal = {Journal of Computational Physics},
    author = {Greencard, L and Rokhlin, V},
    pages = {315--348},
    volume = {73}
}

@article{Cummins2018ADandelion,
    title = {{A separated vortex ring underlies the flight of the dandelion}},
    year = {2018},
    journal = {Nature 2018 562:7727},
    author = {Cummins, Cathal and Seale, Madeleine and Macente, Alice and Certini, Daniele and Mastropaolo, Enrico and Viola, Ignazio Maria and Nakayama, Naomi},
    number = {7727},
    month = {10},
    pages = {414--418},
    volume = {562},
    publisher = {Nature Publishing Group},
    url = {https://www.nature.com/articles/s41586-018-0604-2},
    doi = {10.1038/s41586-018-0604-2},
    issn = {1476-4687},
    pmid = {30333579}
}

@article{Weymouth2014ChaoticCylinder,
    title = {{Chaotic rotation of a towed elliptical cylinder}},
    year = {2014},
    journal = {Journal of Fluid Mechanics},
    author = {Weymouth, G. D.},
    pages = {385--398},
    volume = {743},
    publisher = {Cambridge University Press},
    url = {https://www.cambridge.org/core/journals/journal-of-fluid-mechanics/article/abs/chaotic-rotation-of-a-towed-elliptical-cylinder/254596F5E2FB45E6A456D819827DB57B},
    doi = {10.1017/JFM.2014.42},
    issn = {0022-1120},
    arxivId = {1309.5858}
}

@article{Weymouth2022Data-drivenProjection,
    title = {{Data-driven Multi-Grid solver for accelerated pressure projection}},
    year = {2022},
    journal = {Computers {\&} Fluids},
    author = {Weymouth, Gabriel D.},
    month = {10},
    pages = {105620},
    volume = {246},
    publisher = {Pergamon},
    doi = {10.1016/J.COMPFLUID.2022.105620},
    issn = {0045-7930},
    arxivId = {2110.11029}
}

@article{Rodriguez2011Direct3700,
    title = {{Direct numerical simulation of the flow over a sphere at Re = 3700}},
    year = {2011},
    journal = {Journal of Fluid Mechanics},
    author = {Rodriguez, Ivette and Borell, Ricard and Lehmkuhl, Oriol and Perez Segarra, Carlos D. and Oliva, Assensi},
    month = {7},
    pages = {263--287},
    volume = {679},
    publisher = {Cambridge University Press},
    doi = {10.1017/jfm.2011.136},
    issn = {14697645}
}

@article{Auguste2013FallingDisks,
    title = {{Falling styles of disks}},
    year = {2013},
    journal = {Journal of Fluid Mechanics},
    author = {Auguste, Franck and Magnaudet, Jacques and Fabre, David},
    pages = {388--405},
    volume = {719},
    publisher = {Cambridge University Press},
    url = {https://www.cambridge.org/core/journals/journal-of-fluid-mechanics/article/abs/falling-styles-of-disks/AF4C39F2CEB6C01F069C6049AC352D72},
    doi = {10.1017/JFM.2012.602},
    issn = {0022-1120}
}

@article{Morton1984GeophysicalTy,
    title = {{Geophysical {\&} Astrophysical Fluid Dynamics The generation and decay of vorticity The Generation and Decay of Vort ici ty}},
    year = {1984},
    journal = {Geophys. Asfrophys. Fluid Dynamics},
    author = {Morton, B R},
    pages = {277--308},
    volume = {28},
    publisher = {Gordon and Breach Science Publishers Inc},
    url = {https://doi.org/10.1080/03091928408230368},
    doi = {10.1080/03091928408230368},
    issn = {1029-0419}
}

@article{Koumoutsakos1995High-resolutionMethods,
    title = {{High-resolution simulations of the flow around an impulsively started cylinder using vortex methods}},
    year = {1995},
    journal = {J. Fluid Mech},
    author = {Koumoutsakos, P and Leonard, A},
    pages = {1--38},
    volume = {296}
}

@article{Vincent2016HolesCoins,
    title = {{Holes stabilize freely falling coins}},
    year = {2016},
    journal = {Journal of Fluid Mechanics},
    author = {Vincent, Lionel and Shambaugh, W. Scott and Kanso, Eva},
    month = {8},
    pages = {250--259},
    volume = {801},
    publisher = {Cambridge University Press},
    url = {https://www.cambridge.org/core/journals/journal-of-fluid-mechanics/article/abs/holes-stabilize-freely-falling-coins/944D40B0B3FF9D5EF0A3B796D2D27F1B},
    doi = {10.1017/JFM.2016.432},
    issn = {0022-1120}
}

@article{Carleton2024KirigamiFlow,
    title = {{Kirigami sheets in fluid flow}},
    year = {2024},
    journal = {Extreme Mechanics Letters},
    author = {Carleton, A. G. and Modarres-Sadeghi, Y.},
    month = {9},
    pages = {102198},
    volume = {71},
    publisher = {Elsevier},
    url = {https://www.sciencedirect.com/science/article/abs/pii/S2352431624000786},
    doi = {10.1016/J.EML.2024.102198},
    issn = {2352-4316}
}

@article{Lamoureux2025Kirigami-inspiredReconfiguration,
    title = {{Kirigami-inspired parachutes with programmable reconfiguration}},
    year = {2025},
    journal = {Nature 2025 646:8083},
    author = {Lamoureux, Danick and Fillion, Jérémi and Ramananarivo, Sophie and Gosselin, Frédérick P. and Melancon, David},
    number = {8083},
    month = {10},
    pages = {88--94},
    volume = {646},
    publisher = {Nature Publishing Group},
    url = {https://www.nature.com/articles/s41586-025-09515-9},
    doi = {10.1038/s41586-025-09515-9},
    issn = {14764687},
    pmid = {41034530}
}

@article{Bose2025PorousIncidence,
    title = {{Porous plates at incidence}},
    year = {2025},
    journal = {Theoretical and Computational Fluid Dynamics 2025 39:2},
    author = {Bose, Chandan and Bruce, Callum and Viola, Ignazio Maria},
    number = {2},
    month = {2},
    pages = {19-},
    volume = {39},
    publisher = {Springer},
    url = {https://link.springer.com/article/10.1007/s00162-025-00740-6},
    doi = {10.1007/S00162-025-00740-6},
    issn = {1432-2250},
    arxivId = {2303.13296}
}

@article{Fong2009TheMethod,
    title = {{The black-box fast multipole method}},
    year = {2009},
    journal = {Journal of Computational Physics},
    author = {Fong, William and Darve, Eric},
    number = {23},
    month = {12},
    pages = {8712--8725},
    volume = {228},
    publisher = {Academic Press Inc.},
    doi = {10.1016/j.jcp.2009.08.031},
    issn = {10902716}
}

@article{Collins1973TheCylinder,
    title = {{The initial flow past an impulsively started circular cylinder}},
    year = {1973},
    journal = {Quarterly Journal of Mechanics and Applied Mathematics},
    author = {Collins, W. M. and Dennis, S. C.R.},
    number = {1},
    month = {2},
    pages = {53--75},
    volume = {26},
    doi = {10.1093/QJMAM/26.1.53},
    issn = {00335614}
}

@article{Andersen2005UnsteadyPlates,
    title = {{Unsteady aerodynamics of fluttering and tumbling plates}},
    year = {2005},
    journal = {Journal of Fluid Mechanics},
    author = {Andersen, A. and Pesavento, U. and Wang, Z. Jane},
    month = {10},
    pages = {65--90},
    volume = {541},
    publisher = {Cambridge University Press},
    url = {https://www.cambridge.org/core/journals/journal-of-fluid-mechanics/article/abs/unsteady-aerodynamics-of-fluttering-and-tumbling-plates/1FC26E817B4D5B91714C9875F070CF8D},
    doi = {10.1017/S002211200500594X},
    issn = {1469-7645}
}

@article{Yun2006VorticalNumbers,
    title = {{Vortical structures behind a sphere at subcritical Reynolds numbers}},
    year = {2006},
    journal = {Physics of Fluids},
    author = {Yun, Giwoong and Kim, Dongjoo and Choi, Haecheon},
    number = {1},
    volume = {18},
    publisher = {American Institute of Physics Inc.},
    doi = {10.1063/1.2166454},
    issn = {10706631}
}

@article{Weymouth2025WaterLily.jl:Bodies,
    title = {{WaterLily.jl: A differentiable and backend-agnostic Julia solver for incompressible viscous flow around dynamic bodies}},
    year = {2025},
    journal = {Computer Physics Communications},
    author = {Weymouth, Gabriel D. and Font, Bernat},
    month = {10},
    pages = {109748},
    volume = {315},
    publisher = {North-Holland},
    url = {https://linkinghub.elsevier.com/retrieve/pii/S0010465525002504},
    doi = {10.1016/J.CPC.2025.109748},
    issn = {0010-4655}
}

\end{document}